\documentclass[12pt]{article}

\usepackage{newtxtext,newtxmath}
\usepackage{graphicx}
\usepackage{comment}
\usepackage[letterpaper,margin=1in]{geometry}

\renewenvironment{abstract}
	{\quotation}
	{\endquotation}
\date{}

\makeatletter
\renewcommand{\fnum@figure}{\textbf{Figure \thefigure}}
\renewcommand{\fnum@table}{\textbf{Table \thetable}}
\makeatother

\usepackage{scicite}

\usepackage{url}

\def\scititle{
	Giant enhancement of exciton diffusion near an electronic Mott insulator
}

\title{\bfseries \boldmath \scititle}

\author{
    Pranshoo Upadhyay$^{1,\ast}$,
    Daniel G. Su\'arez-Forero$^{1,2,\ast,\dagger}$,
    Tsung-Sheng Huang$^{1,\ast}$, \and
    Mahmoud Jalali Mehrabad$^{1}$,
    Beini Gao$^{1}$,
    Supratik Sarkar$^{1}$,
    Deric Session$^{1}$, \and
    Kenji Watanabe$^{3}$,
    Takashi Taniguchi$^{3}$,
    You Zhou$^{4,5}$, \and
    Michael Knap$^{6,7}$,
     Mohammad Hafezi$^{1,\ddagger}$\and
	\small$^{1}$Joint Quantum Institute, University of Maryland, College Park, MD 20742, USA.\and
    \small$^{2}$Department of Quantum Matter Physics, University of Geneva, Geneva, Switzerland.\and
    \small$^{3}$Research Center for Materials Nanoarchitectonics, National Institute for Materials Science,\and \small 1-1 Namiki, Tsukuba 305-0044, Japan.\and
    \small$^{4}$Department of Materials Science and Engineering, University of Maryland,\and \small College Park, MD 20742, USA.\and
    \small$^{5}$Maryland Quantum Materials Center, College Park, Maryland 20742, USA.\and
    \small$^{6}$Technical University of Munich, TUM School of Natural Sciences, \and \small Physics Department, 85748 Garching, Germany.\and
    \small$^{7}$Munich Center for Quantum Science and Technology (MCQST), \and \small Schellingstr. 4, 80799 M{\"u}nchen, Germany.\and
	\small Corresponding authors. Email: $^\dagger$ suarez@umbc.edu, $^\ddagger$ hafezi@umd.edu \and
	\small$^\ast$These authors contributed equally to this work.
}

\begin{document} 

\maketitle
\begin{abstract} \bfseries \boldmath
Bose-Fermi mixtures naturally appear in various physical systems. In semiconductor heterostructures, such mixtures can be realized, with bosons as excitons and fermions as dopant charges. However, the complexity of these hybrid systems challenges the comprehension of the mechanisms that determine physical properties such as mobility. In this study, we investigate interlayer exciton diffusion in an H-stacked WSe$_2$/WS$_2$  heterobilayer. Our measurements are performed in the ultra-low exciton density regime at low temperatures to examine how the presence of charges affects exciton mobility. Remarkably, for charge doping near the Mott insulator phase, we observe a giant enhancement of exciton diffusion of three orders of magnitude compared to charge neutrality. We attribute this observation to mobile valence holes, which experience a suppressed moir\'e potential due to the electronic charge order in the conduction band, and recombine with any conduction electron in a non-monogamous manner. This new mechanism emerges for sufficiently large fillings in the vicinity of correlated generalized Wigner crystal and Mott insulating states. Our results demonstrate the potential to characterize correlated electron states through exciton diffusion and provide insights into the rich interplay of bosons and fermions in semiconductor heterostructures. 
\end{abstract}

\noindent
Layered transition metal dichalcogenides (TMDs) have become an interesting platform to study collective emerging electronic phenomena, including Mott insulators \cite{SimHubb2020, MottAtac2020,shimazaki2021optical}, generalized Wigner crystals \cite{Fractstat2020, MottandWig2020}, density waves \cite{EDW2023}, fractional Chern insulators \cite{FQAH2023, FCI2023}, quasi-exciton condensation \cite{EI2021Mak}, superconductivity \cite{SCTWSe22024,Guo2024-zb}, and kinetic ferromagnetism \cite{KineticM2023}. One remarkable direction is to use optical %excitation of 
excitons %and exploit them 
to probe various electronic orders \cite{du2023moire}. Since excitons can be generally considered as bosons, hybrid electron-exciton systems provide a natural platform for investigating Bose-Fermi mixtures. Bose-Fermi mixtures are ubiquitous in many-body physics, from solutions of fermionic $^3$He in bosonic superfluid $^4$He \cite{Ebner1971-yu}, and quark-meson models in QCD physics \cite{Schaefer2005-md,achenbach2024present}, to ultracold atoms \cite{gunter2006bose,schreck2001sympathetic,hu2016bose,delehaye2014mixture}.

Recent experimental implementations of Bose-Fermi mixtures in TMDs \cite{Sufei2023, Gao2024, xiong2023correlated,Dipoleladd2023,Ma2021-yu,Zhang2022-rs} have motivated the search for exotic physics such as tunable particle interactions \cite{kuhlenkamp2022tunable,schwartz2021electrically}, and topological superconductivity \cite{zerba2024realizing,Andrei2021-ur}. A key challenge is to understand how excitons behave in correlated electronic environments. Measuring the transport of excitons immersed in an electron gas has the potential to address this challenge. This technique has been utilized earlier in quantum well platforms \cite{Sanvitto2001-kr,Pulizzi2003-mx} and later in TMD systems \cite{DiffRev2023,DiffTrap2020,DistinctphasesDiff2021,jin2018imaging,Jauregui2019-zk, DiffDipole2022, Difftwist2020, PRL2024NegDiff, ElectricDiffDipol2023,rossi2024anomalous} to probe the physics of excitons when embedded in a Fermi sea. Adding correlated phenomena into this picture has opened unprecedented perspectives to the field, bringing together emerging collective states and light-matter interaction. Here, we explore the limit in which a very dilute concentration of bosons (excitons) diffuses in a 2D fermionic (electronic) landscape, in the absence of exciton-exciton interactions, to investigate how the rich phases of the fermionic many-body system affect the dynamics of bosons.

Our system consists of a WSe$_2$/WS$_2$ moir\'e heterostructure where we employ space- and time-resolved techniques to study diffusing interlayer excitons (IXs) immersed in a 2D fermionic electron gas. Our measurements reveal that the exciton diffusion coefficient is highly sensitive to changes in the electronic filling of the system. We analyze exciton dynamics within the various exotic electronic states that are realized by our system, uncovering a rich landscape dominated by polarons, generalized Wigner crystals, and a Mott-insulating state. Depending on the electronic state, we observe dramatic variations in the mobility of the diffusing species, with changes up to three orders of magnitude. Our results challenge the common assumption that electrons and holes forming IXs always move together in a monogamous manner, particularly for fillings slightly below the Mott-insulating regime. With the help of an effective model Hamiltonian, an interplay of two different channels for the diffusion of excitonic species is demonstrated. These findings present a novel optical approach for exploring complex quantum states in condensed matter systems.

\section{Physical system}
\begin{figure*}
\centering
\includegraphics[width=1\linewidth]{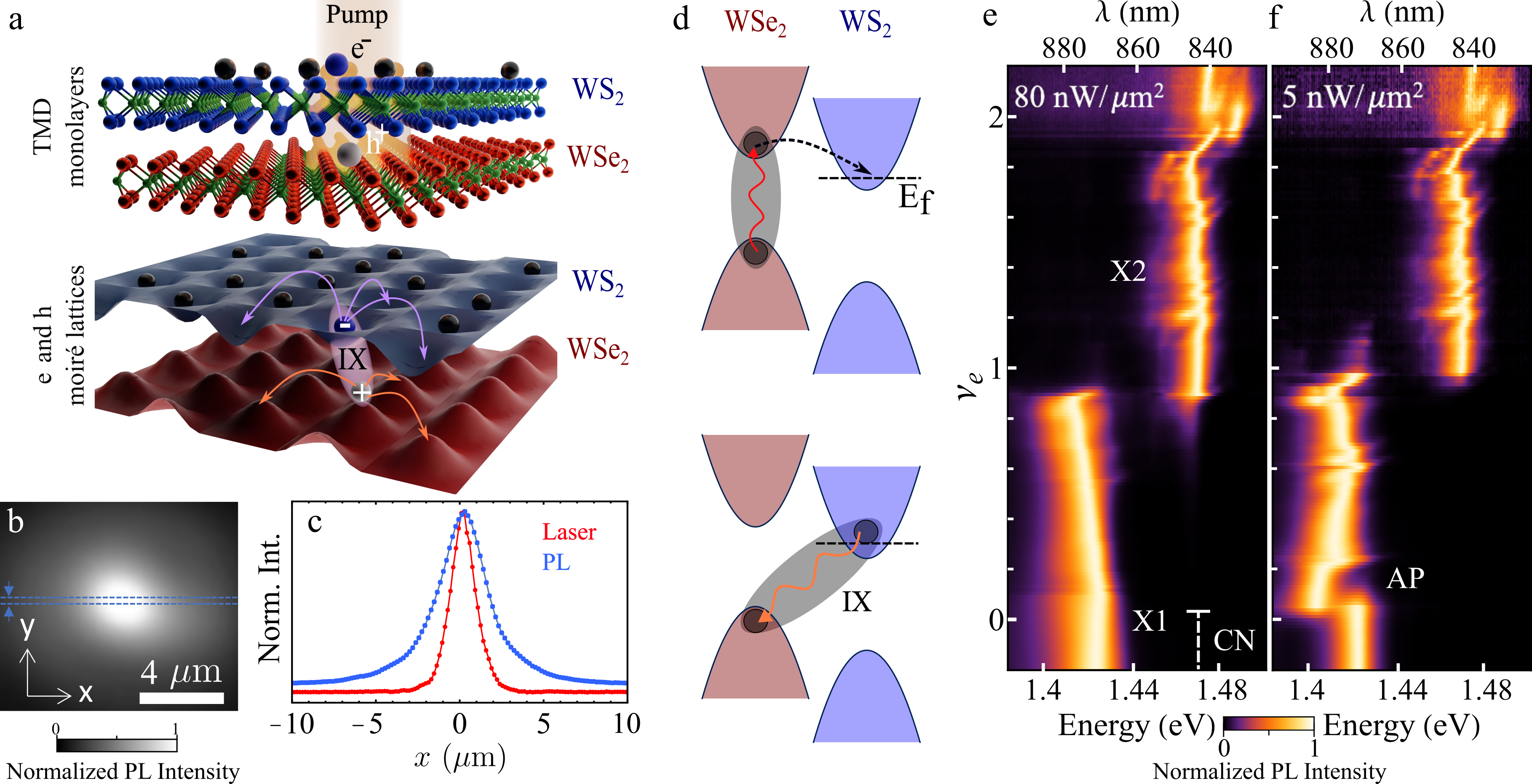}
    \caption{\textbf{Description of the experimental scheme and characterization of the physical system.} (a) Schematic of the diffusion setup for the extraction of IX hopping strength in the moir\'e system in a doped environment. (b) Image of the IX PL when excited with a diffraction-limited laser spot. (c) The spatial variation of PL intensity (blue) along the x-axis within the region indicated by the blue dashed line in (b) is compared to that of the laser spot (red). The difference between the laser profile and the PL profile originates in the diffusion of the excitonic population. (d) IX creation in the bilayer sample. A resonant laser creates a population of excitons in the WSe$_2$ layer (upper panel). Due to the band alignment, the electron tunnels from the WSe$_2$ to the WS$_2$ and forms an IX that diffuses and recombines radiatively. Normalized PL spectrum of IX at 3.5 K as a function of electronic filling factor, $\nu_{\rm{e}}$ at (e) moderate (80 nW/$\mu m^2$) and (f) ultra-low (5 nW/$\mu m^2$) excitation pump intensity. The blue-shift at $\nu_{\rm{e}} \sim\! 1$ indicates the Mott gap ($U_{\rm{ex}-\rm{e}} \sim\!46$ meV). The polaron gap ($\sim\! 16$ meV) is observed only at ultra-low pump intensities.}
    \label{fig1}
\end{figure*}

The experimental setup and moir\'e system are depicted in Fig.~\ref{fig1}a. The system consists of an H-stacked WS$_2$/WSe$_2$ bilayer heterostructure hosting a triangular moir\'e lattice (see Supplementary Material). Optically exciting this system creates IX that form between spatially separated electrons and holes residing in different moir\'e registries \cite{ElectronspreadH2023}. The nature and dynamics of IX is determined by the state of the electron gas. The spatially resolved photoluminescence (PL) emission, which is broader than the diffraction-limited optical excitation, encodes the information of IX diffusion (Fig.~\ref{fig1}b-c). The excitation pump is resonant with the WSe$_2$ intralayer exciton in all our measurements (upper panel of Fig.~\ref{fig1}d). Upon ultra-fast electron transfer (in the order of femtoseconds \cite{Chargetrans2014}), the IX form, diffuse, and optically recombine (lower panel of Fig.~\ref{fig1}d). As mentioned earlier, in this work we focus on the ultra-low exciton density regime where exciton-exciton interactions are negligible. Hence, the system can be treated as individual bosons moving within a gas of fermions. 

To identify this regime, we look at the collected PL spectrum of IX as a function of the electron filling ($\nu_e$) for different pump intensities. Details on the calibration of $\nu_{\rm{e}}$ are provided in Supplementary Note 1. Figure \ref{fig1}e-f displays the normalized PL spectra for two pump intensities (80 nW/$\mu$m$^2$ and 5 nW/$\mu$m$^2$). For both exciton densities, the charge neutral (CN) region is dominated by a low-energy IX (X1) which transitions into a high-energy exciton (X2) at $\nu_e\!\sim\!1$. The transition is benchmarked by a $\sim\!46$ meV gap in the PL emission, and it has been recently demonstrated to arise from the strong on-site exciton-electron repulsions owing to the formation of an electronic Mott-insulating state at $\nu_e\!\approx 1$ \cite{Gao2024,xiong2023correlated,Sufei2023}. Although the system's response in the Mott insulating state is similar for both pump intensities, there is a clear difference at low electron densities: in the ultra-low exciton density case (panel f) an additional gap of $\sim\!16$ meV is observed near $\nu_e=0$. This redshift corresponds to the transition from X1 to attractive polaron (AP), as corroborated by the diffusion measurements discussed later. The AP formation is facilitated in such an H-stacked system where the hole can bind with multiple nearest neighboring electrons \cite{ElectronspreadH2023}; more details regarding the AP formation can be found in Supplementary Note 2. Understanding the precise nature of the AP in the presence of a lattice is an open problem and will be an interesting future research direction both experimentally and theoretically. Here, we identify the AP with the dressed charge complex comprised of X1 and dopant charges \cite{liu2021signatures, wang2021moire}. It is important to note that AP formation occurs in both cases when the system is doped, however, a complete transition into AP can only be observed when the ratio of X1 to electron density is small; in other words, when each exciton can be dressed by surrounding electrons. This observation is validated by even lower pump intensity ($<$ 5 nW/$\mu m^2$) measurements, as shown in Supplementary Note 3. All the presented data has been taken in this low-pump intensity regime unless stated otherwise. 

\begin{figure*}
    \centering
    \includegraphics[width=\linewidth]{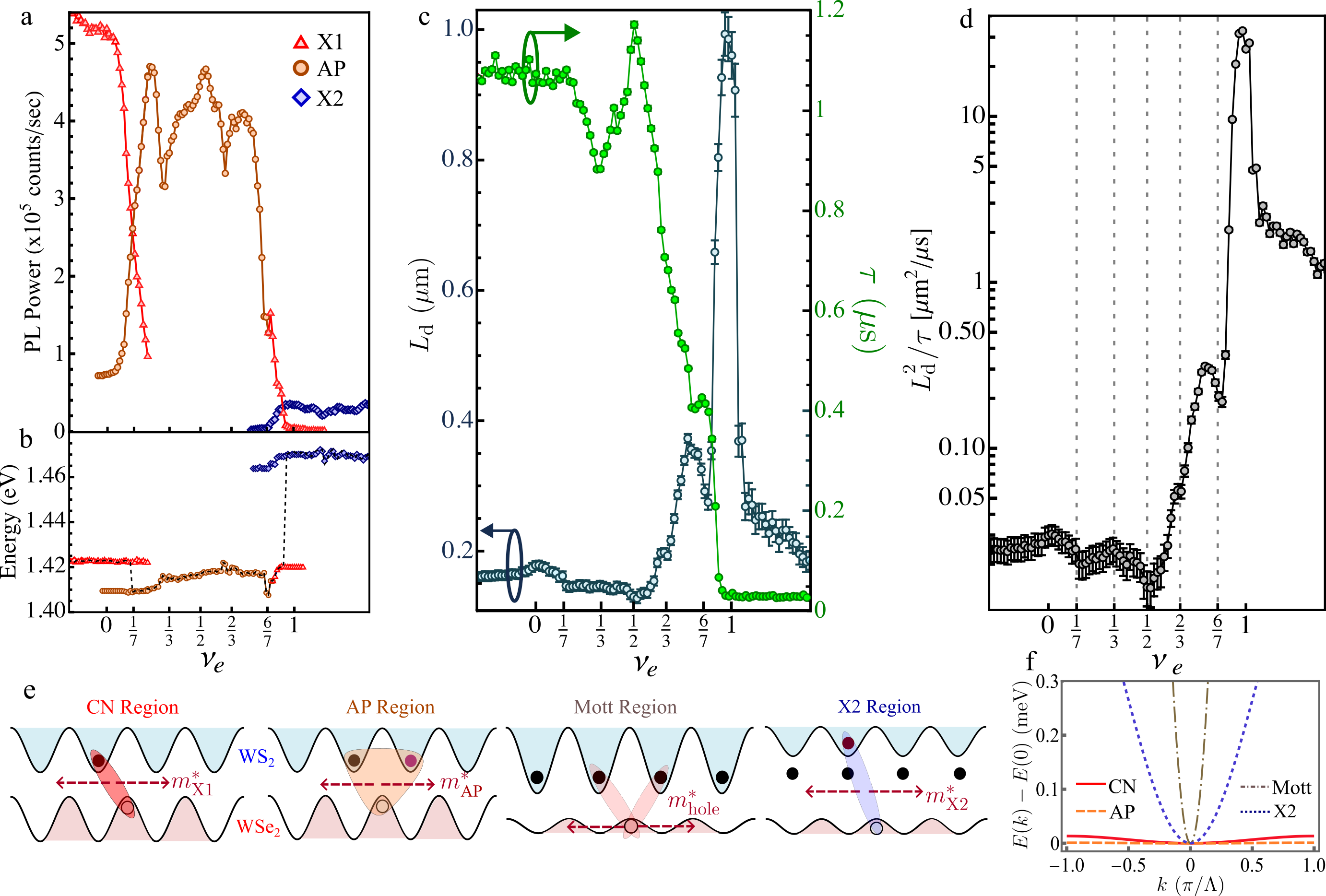}
    \caption{\textbf{Space- and time-resolved measurements of interlayer quasiparticle dynamics in an electron-doped lattice.} (a) Integrated PL of X1, AP, and X2 as a function of doping aids in identifying different regimes of fermionic filling, including the onset of the AP and the Mott-insulating region. (b) Peak energy tracking as a function of doping. (c) Diffusion length ($L_{\rm{d}}$ shown in blue) and IX lifetime ($\tau$ shown in green) as a function of electron doping. Both quantities strongly vary around fractional and integer fillings.  (d) Diffusivity, $D\!=\!\frac{L_{\rm{d}}^{2}}{\tau}$, demonstrates the variation of the tunneling rate of interlayer quasiparticles. Remarkably, more than three orders of magnitude enhancement in $D$ is observed near the Mott-insulating state. All measurements are collected at a temperature of 3.5 K. (e) Four fermionic regions highlighting variation of tunneling species are demonstrated. The shifted moir\'e lattice for holes and electrons represents the H-stacked system. AP being heavier than X1 results in lower $D$. For the case of $\nu_{\rm{e}}\sim\! 1$, X1 does not tunnel as a composite particle rather the hole moves in a non-monogamous way.  In this configuration, the effective potential for holes in the Mott region is reduced. For even larger fillings $\nu_{\rm{e}}\gtrsim 1$, X2 tunnels in an effectively reduced moir\'e potential. (f) Energy dispersion of the diffusing particles, shown in (e), calculated for these regions (color encoded) indicate the variation of the effective mass ($m^*$) in the system.}
    \label{fig2}
\end{figure*}

In this low-pump intensity regime, we identify four distinct spectral features as a function of electron doping as shown in Fig.~\ref{fig1}f. The extracted intensity and energy values are shown in Fig.~\ref{fig2}a-b, and they can be distinguished as follows:
\begin{enumerate}
    \item Charge-neutral region: dominated by X1 emission,
    \item Moderate-doping region ($1/7 \lesssim \nu_{\rm{e}} \lesssim 6/7 $): Emission from AP appears as soon as doping begins. However, for filling factors within this range, AP dominates the PL emission. Exception occurs at certain fractional fillings (e.g. 1/3, 2/3, and 6/7) that host correlated states.
    \item Mott-insulating region ($6/7 \gtrsim\nu_{\rm{e}} \gtrsim 1$): dominated by X1. The strong electron localization inhibits the formation of APs in this region.
    \item High-doping region ($\nu_e\! \gtrsim\!1$): dominated by X2 due to double occupancy of moir\'e sites (doublon-hole pair).
\end{enumerate}
We focus on the electron-doped H-stack case due to its pronounced features, including a well-defined AP region and a significant Mott gap. Supplementary Note 4 provides additional data on the extended doping regimes in both H- and R-stacked devices. Having identified the four different doping regimes, we now measure the diffusion dynamics of the quasiparticles of each regime and their dependence on the density of the 2D electron gas.

\section{Exciton diffusion mechanisms}

We extract the diffusion length from the spatially-resolved PL emission by employing a Gaussian fitting routine and analyze it as a function of $\nu_e$. Considering that the measured PL profile corresponds to a convolution between quasiparticle diffusion and laser intensity distribution, the characteristic diffusion length is given by $L_{\rm{d}} = \sqrt{\sigma_{\rm{X}}^2-\sigma_{\rm{L}}^2}$, where $\sigma_{\rm{X}}$ and $\sigma_{\rm{L}}$ correspond to the half-widths of the PL and laser profiles, respectively. The detailed calculation of extracting $L_{\rm{d}}$ in the steady-state condition is presented in Supplementary Note 5. Furthermore, as the emission wavelength across doping regimes considerably varies, we check our setup to discard any chromatic aberrations, refer to Supplementary Note 6. After these careful considerations, the obtained values for the diffusion length are displayed as blue markers in Fig.~\ref{fig2}c. In the CN region, the diffusion length is less than 200 nm; over an order of magnitude lower than previously reported measurements in structures without a moir\'e lattice  \cite{NegligDiffACS2021, Moireimpedes2020}. Such a suppression in diffusion length indicates that the IX kinetic energy is quenched by the presence of a strong moir\'e potential in our system. Further suppression of $L_{\rm{d}}$ is observed upon doping the system, i.e. in the region dominated by AP. Remarkably, a significant enhancement of $L_{\rm{d}}$ is noticed as the system is further doped to $\nu_{\rm{e}}\!\sim\!1$, followed by a significant drop in the regime dominated by X2. For further clarity, spectrally resolved diffusion is shown in Supplementary Note 7. 

A complete analysis of the IX mobility requires the measurement of its lifetime ($\tau$), as the modulations in diffusion can originate either from the tunneling rate variation or the quasiparticle's lifetime. We measure this quantity by using a femtosecond pulsed laser with a repetition rate of $\sim\!500$ kHz while the signal is collected in a superconducting single photon detector (see Methods). The observed lifetime as a function of $\nu_{\rm{e}}$ is shown with green markers in Fig.~\ref{fig2}c. We notice a trend of lifetime reduction with doping, and modulations at some fractional fillings. In agreement with previous reports \cite{Dipoleladd2023}, $\tau$ decreases by over an order of magnitude, from $\sim\!1$ $\mu$s in the CN region to $\sim\!30$ ns in the Mott region. The reduction in $\tau$ and enhancement of $L_{\rm{d}}$ in the Mott region suggests the presence of highly mobile particles with low effective mass. 

Assuming that the dynamics is diffusive, we can use the diffusion coefficient ($D = L_{\rm{d}}^2/\tau$) to study different regimes. We observe a remarkable three-orders of magnitude enhancement of the mobility at the Mott-insulating region, see Fig.~\ref{fig2}d. This result is verified in different devices and two independent setups (see Supplementary Notes 4 and 8). This is an intriguing behavior since the electron mobility is expected to be suppressed at densities below the Mott insulating regime.
 To comprehend this anomalous behavior, we analyze how the changes in $\nu_{\rm{e}}$ lead to the renormalization of the diffusing particle's effective mass ($m^*$). This effect originates from the Coulomb potential ($V_{\rm{Coul}}$) exerted by the electron gas on the diffusing particles. The addition of $V_{\rm{Coul}}$ and the moir\'e potential, together with changes in the mass of the diffusing particles determine the large variations in $m^*$. An effective model to describe this behavior \cite{lagoin2021key} and the extracted values of $m^*$ for each doping region are presented in Supplementary Note 9.

\begin{figure*}
    \centering
    \includegraphics[width=1\linewidth]{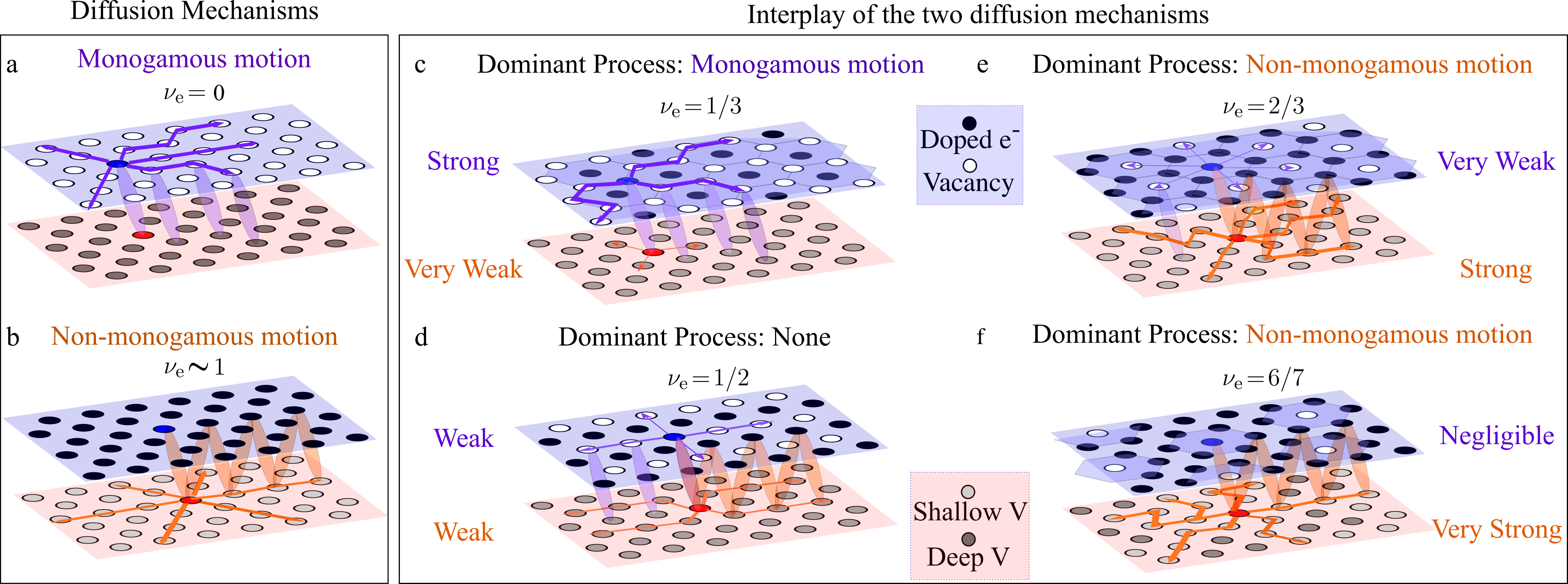}
    \caption{\textbf{Illustration of the exciton diffusion mechanisms.} The two diffusion channels that determine the motion of optically created electron (blue circle) and hole (red circle) pairs are displayed in panels (a) and (b). Black and white circles in the electron layer indicate electronically filled and vacant sites, respectively. Light-gray and dark-gray circles in the hole layer indicate shallow and deep effective hole potentials. (a) At $\nu_{\rm{e}}\!=\!0$, only monogamous IX motion is possible. (b) For $\nu_{\rm{e}}\!\sim\!1$, the only channel is the non-monogamous hole hopping. The interplay of these two channels is represented for the following filling fractions: (c) $\nu_{\rm{e}}\!=\!1/3$, (d) $\nu_{\rm{e}}\!=\!1/2$, (e) $\nu_{\rm{e}}\!=\!2/3$, and (f) $\nu_{\rm{e}}\!=\!6/7$. Available NN vacant sites enhance monogamous motion, while NN electron-doped sites enhance non-monogamous hole hopping. The trend reverses at $\nu_{\rm{e}}\!=\!1/2$, where both processes are effectively suppressed.}
    \label{fig3}
\end{figure*}

The four physical situations depicted in Fig.~\ref{fig2}e describe how the electrons affect $D$. With the mentioned model, we estimate the energy dispersion of the diffusing particles for each case (Fig.~\ref{fig2}f). For the CN region, as $V_{\rm{Coul}} \approx 0$, the electron and hole experience a similar potential landscape in their respective layers, favoring their motion together as a bound particle. As expected for moir\'e-trapped excitons, we observe a diffusion coefficient several orders of magnitude lower than the values reported for IX in the absence of moir\'e potential \cite{Moireimpedes2020}. In the AP region, where $V_{\rm{Coul}}$ remains comparatively negligible, we detect an increase in the effective mass $m^*_{\rm{AP}}$. The formation of heavier quasiparticles resulting from the dopant electrons dressing the injected optical excitons leads to the observed reduction in mobility. The decreasing trend in $D$ reverses at $\nu_{\rm{e}}= 1/2$. This phenomenon is discussed in more detail in the next section.

The dramatic three-orders-of-magnitude increase in mobility near the Mott region can be attributed to the stronger effective potential in the electron layer and the suppressed potential in the hole layer, as depicted in Fig.~\ref{fig2}e. This significant variation in effective potential results from the crystallization of electrons, which generates $V_{\rm{Coul}}$ with the same period and phase as the electron moiré lattice but out of phase with the hole moiré lattice as holes occupy different atomic registry. Additionally, due to the absence of any vacancies in the electron layer, the hole in the other layer is free to hop into any moir\'e site. Hence, in the Mott region, the hole is not bound to a specific electron and can recombine with any electron in the lattice in a non-monogamous way. This non-monogamous hopping, combined with the reduced effective potential depth, leads to a decrease in $m^*$, accounting for the orders-of-magnitude enhancement in $D$. However, beyond $\nu_{\rm{e}}= 1$, the excited electrons become mobile again, allowing the electron and hole to move together, in contrast to the previous non-monogamous hole dynamics. This is pictorially shown in the last panel of Fig.~\ref{fig2}e, where an additional electron on top of the electron lattice and a hole in the other layer form a diffusing exciton. Interestingly, in this X2 region, both the hole and the electron encounter shallower potentials due to the out-of-phase $V_{\rm{Coul}}$ in each layer. This shallower potential results in orders-of-magnitude increased mobility of X2 compared to X1. However, the monogamous motion manifests as a reduction in mobility compared to the holes in the Mott region. This discussion elucidates the observed strong variation in $D$ across the four different fermionic regions.

\section{Diffusion at generalized Wigner crystal states}

The remarkable sensitivity of the mobility to the fermionic state of the system is further manifested at certain fractional fillings that host generalized Wigner crystals due to the long-range repulsive interactions. Before focusing on these specific filling factors, we reiterate that the general trend of $D$ is reversed at $\nu_{\rm{e}}\!=\!1/2$. This observation can be understood by analyzing the occupation of the nearest neighbor (NN) sites of the injected IX below and above $\nu_{\rm{e}}\!=\!1/2$. Figure~\ref{fig3} illustrates how, in the two regimes, the NN filling determines the interplay between the two diffusion mechanisms. At $\nu_{\rm{e}}\!=\!0$, where all the NN sites are vacant, only monogamous motion can take place (Fig.~\ref{fig3}a), while for $\nu_{\rm{e}}\!\sim\!1$, non-monogamous motion exclusively determines the exciton mobility due to the complete occupation of the moir\'e sites (Fig.~\ref{fig3}b). For any other filling, the system experiences an interplay between the two diffusion mechanisms.

To elucidate, we make a direct comparison of the available diffusion channels at  $\nu_{\rm{e}}\!=\!1/3$ and $\nu_{\rm{e}}\!=\!2/3$. The monogamous motion of the exciton, allowed for $\nu_{\rm{e}}\!=\!1/3$, is blocked by the filled NN sites in the electron layer at $\nu_{\rm{e}}\!=\!2/3$ (Fig.~\ref{fig3}c and e). Hence, one would expect lower mobility at $\nu_{\rm{e}}\!=\!2/3$ compared to $\nu_{\rm{e}}\!=\!1/3$. However, the experimental results (Fig.~\ref{fig2}d) show a local minimum but an overall increase in $D$ at $\nu_{\rm{e}}\!=\!2/3$ compared to $\nu_{\rm{e}}\!=\!1/3$. To understand this, we note that while the occupied NN sites hinder the monogamous motion of electron and hole, they simultaneously create non-monogamous channels that facilitate the diffusion of holes, as shown by the arrows in the bottom layer of Fig.~\ref{fig3}e. Therefore, the increasing trend in $D$ is a consequence of the dominant non-monogamous hole diffusion. This effect gets stronger at $\nu_{\rm{e}} \!=\! 6/7$ as the occupation of both NN and next-nearest neighbor (NNN) sites (Fig.~\ref{fig3}f) increases the number of diffusion channels in the hole layer. It should be noted that the local minima in $D$ at $\nu_{\rm{e}}\!=\!2/3$ and $6/7$ result from the inhibition of the monogamous motion of IXs because the electrons' wavefunctions are more localized.

The interplay of the two discussed diffusion mechanisms determines the non-trivial dependence of $D$ in the explored range of $\nu_{\rm{e}}$. In this picture, the dominant mechanism changes at $\nu_{\rm{e}}= 1/2$. The monogamous motion of quasiparticles (X1 or AP) determines the mobility in the region 0 $< \nu_{\rm{e}}<$ 1/2, while it is the non-monogamous hole diffusion for 1/2 $< \nu_{\rm{e}}<$ 1. $\nu_{\rm{e}}= 1/2$ (Fig.~\ref{fig3}d) acts as the turning point, exhibiting the lowest mobility as both processes are effectively suppressed.

\begin{figure}
    \centering
    \includegraphics[width=0.7\columnwidth]{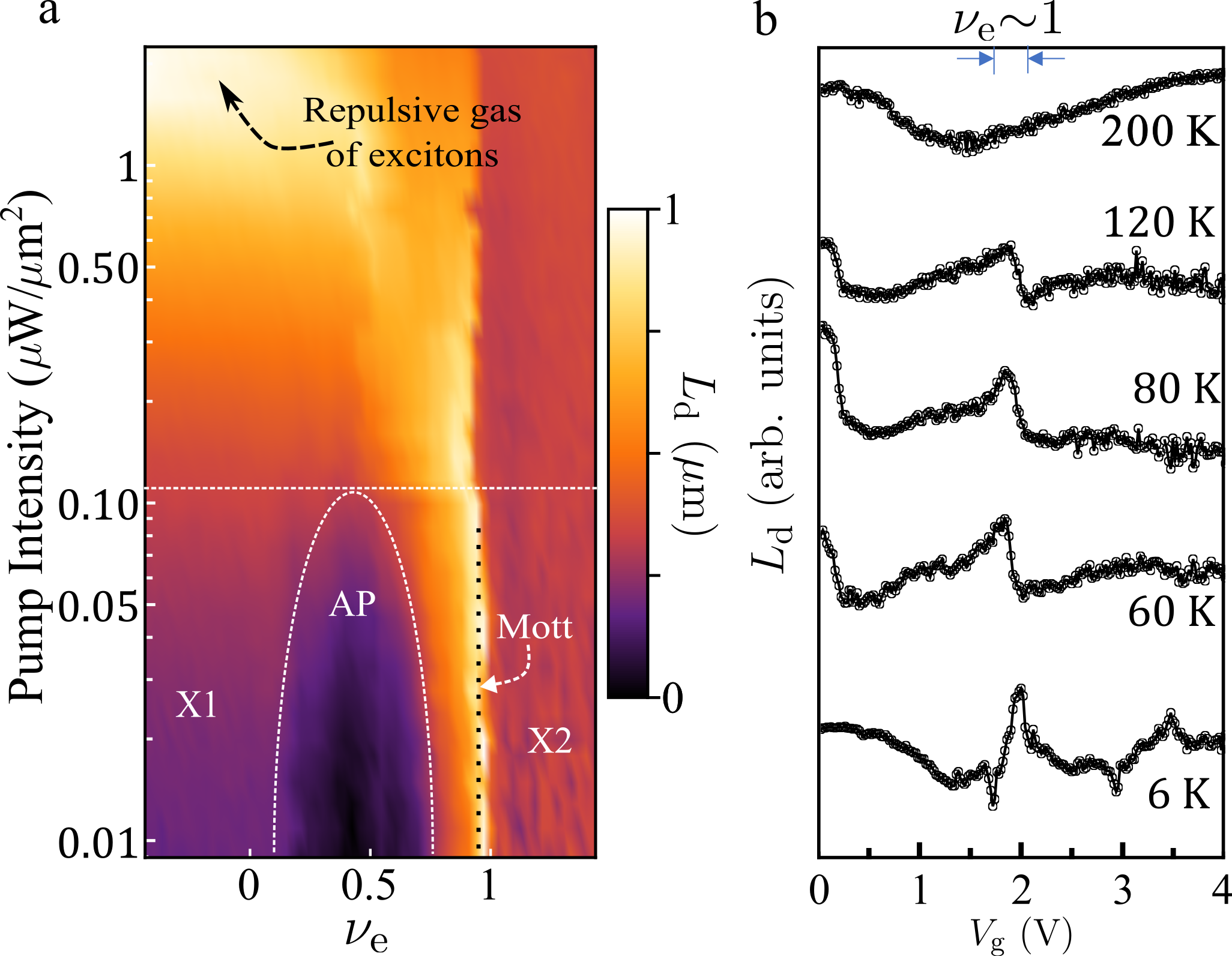}
    \caption{\textbf{Pump intensity and temperature dependence of diffusion in the moir\'e lattice.} (a) The pump intensity-dependent variation demonstrates the threshold (horizontal white dashed line) above which IX-IX scattering mechanisms dictate the diffusion dynamics. Diffusion at pump intensities above this threshold is primarily due to annihilation and repulsion mechanisms while at ultra-low pump intensity, the diffusion originates from quasiparticle tunneling between moir\'e sites. The vanishing of AP and Mott features in $L_{\rm{d}}$ for high pump intensities demonstrates this difference. (b) Temperature dependence of $L_{\rm{d}}$ shows the existence of a peak around $\nu_{\rm{e}}\sim\! 1$ for T $<$ 200 K, although other features, including AP and fractional fillings, are lost. No observable feature in diffusion exists above this temperature although the Mott gap exists for the entire range of temperature used in our measurements. This is due to the enhanced phonon scattering channels that dominate the dynamics. For high temperatures, we use $V_{\rm{g}}$ in the horizontal axis because $\nu_{\rm{e}}$ is not well defined.}
    \label{fig4}
\end{figure}

\section{Pump intensity phase map and temperature dependence}

Our physical picture of the diffusion mechanisms is limited to a regime of ultra-low exciton density at cryogenic temperatures. Additional physical processes affecting the dynamics are expected to emerge for higher densities and temperatures. To gain insights into these more complex regimes, we map the dependence of the diffusion length as a function of $\nu_{\rm{e}}$ and pump intensity (Fig.~\ref{fig4}a). We emphasize that the excitation pump intensity cannot be directly related to the exciton density \cite{Gao2024}. This is because the total PL intensity and exciton lifetime are not constant throughout the doping range (Fig.~\ref{fig2}a). However, as illustrated in Fig.~\ref{fig4}a, the enhancement of the mobility near the Mott region and the modifications at the fractional states exist for every pump intensity below 100 nW/$\mu m^2$ (white dashed line). In this ultra-low exciton density regime, the diffusion features are independent of exciton density variations. Beyond the threshold intensity, the sharp peak associated with the Mott insulating region becomes less distinctive and shifts to lower $\nu_{\rm{e}}$, indicating a regime where additional processes such as double exciton occupancies and exciton-exciton scattering become relevant. The monotonic increase in $L_{\rm{d}}$ with pump intensity at charge neutrality (CN) suggests that these scattering processes are enhanced with the exciton density. For pump intensities greater than 1 $\mu$W$/\mu$m$^2$, the system enters a non-linear repulsive regime where exciton-exciton dipole repulsion dominates the diffusion dynamics. Figure \ref{fig4}a highlights these different regimes.

The reduction of mobility due to the formation of AP is more sensitive to exciton density. Hence, the features in this region start to fade away at around 50 nW/$\mu$m$^2$. In this case, the exciton to polaron transition is blurred out due to the larger population of excitons compared to the dopant electrons in the lattice. 

Another crucial parameter for $L_{\rm{d}}$ is the temperature ($T$). Figure \ref{fig4}b shows that the diffusion peak at the Mott insulating state vanishes for $T\!\geq\!200$ K. Although the Mott gap persists over the full measured temperature range (see Supplementary Note 10), no significant dependence of $L_{\rm{d}}$ with $V_{\rm{g}}$ is observed beyond 200 K. This is attributed to increased phonon-assisted scattering mechanisms that dominate the diffusion dynamics at higher temperatures \cite{CHERNIKOV202369}. Moreover, note that at 200 K the inhomogeneous broadening becomes comparable to the Mott gap, and the giant diffusion enhancement disappears. Additionally, the signatures of AP in the voltage regime 1 V $\!\lesssim\! V_{\rm{g}}\!\lesssim$ 2 V are suppressed above 60 K, as the AP cannot form at elevated temperatures due to its low binding energy. 

\section{Outlook} In conclusion, we unveil possible mechanisms affecting the mobility of a dilute exciton population in an electron-doped moir\'e system.  The observed drastic sensitivity of the exciton diffusion to the nature of the underlying electronic state demonstrates the potential of this technique to probe other complex states of matter. The changes in the binding energy and scattering time of the many-body excitonic species in the presence of correlated fermionic states is important to fully understand the correlations and dynamics in such Bose-Fermi mixtures, and remains an open question both for theory and experiments.

In perspective, a broad range of physical effects can be explored by measuring the mobility of embedded particles in van der Waals structures. For example, the nature of exotic phases of matter, including kinetic magnetism \cite{KineticM2023,Yang2024LFM}, exciton condensation \cite{Yao2008-nx}, and anomalous quantum Hall regimes \cite{FQAH2023,FCI2023}, could be studied with this technique. In particular, incorporating polarization resolution and fields into this technique offers the possibility to study fractional Chern insulators \cite{FCI2023}, Mott-moir\'e excitons \cite{Huang2023}, spin polarons \cite{tao2024observation}, and tunable electron-exciton interactions \cite{kuhlenkamp2022tunable,schwartz2021electrically}. Moreover, the microseconds-long exciton lifetimes allow one to conceive time-resolved diffusion experiments, capable of extracting valuable insights into the temporal dynamics of fermionic correlated states.

%%%%%%%%%%%%%%%% ACKNOWLEDGEMENTS %%%%%%%%%%%%%%%

\section*{Acknowledgement}
The authors acknowledge fruitful discussions with Ming Xie, Ajit Srivastava, and Angel Rubio. This research was partially supported by NSF OMA-2120575, AFOSR FA95502010223, ARO W911NF2010232, and Simons Foundation. YZ acknowledges support from the National Science Foundation under Award No. DMR-2145712. M.K. acknowledges support from the Deutsche Forschungsgemeinschaft (DFG, German Research Foundation) under Germany's Excellence Strategy--EXC--2111--390814868 and from the European Research Council (ERC) under the European Unions Horizon 2020 research and innovation programme (Grant Agreement No. 851161).

\section*{Methods}
\subsection*{Device fabrication}
The hBN encapsulated WSe$_2$/WS$_2$ heterostructures were fabricated using a dry-transfer method reported in the literature \cite{Gao2024}. All flakes were exfoliated from bulk crystals onto Si/SiO$_2$ (285 nm) and identified by their optical contrast. The top/bottom gates and TMD contact are made of few-layer graphene. The flakes were picked sequentially with a polymer stamp and released onto a Si/SiO$_2$ (90 nm) substrate. Later, electrodes consisting of 10 nm of chromium and 70 nm of gold were patterned on the substrate. They were fabricated using standard electron-beam lithography techniques and thermal evaporation. The sample was annealed at 300$^\circ$ C for 2 hr.

\subsection*{Optical Measurements}
All the measurements are performed in a dilution refrigerator at 3.5K unless stated otherwise. The sample is excited using a Ti:Sapphire laser tuned at 733 nm, resonant to the WSe$_2$ intralayer exciton, and focused to diffraction limit with an 80$\times$ microscope objective. Spatially resolved images are collected in a CCD camera (Princeton Instruments Blaze HRX) coupled to a spectrograph (Princeton Instruments SP2750). The total magnification of the optical setup is $\sim\!220\times$.
For lifetime measurements, we excite the sample with a 100 fs pulsed Ti:Sapphire laser. We use a pulse picker to achieve a repetition rate low enough to detect the optical decay of the long-lived IX ($\sim\!500$ kHz). In this case, we use a superconducting nanowire single-photon detector and an event timer module to obtain the time-correlated PL signal.

\section*{Competing interests}
The authors declare no competing interests.

\section*{Data availability}
All of the data that support the findings of this study are reported in the main text and Supplementary Material. Source data are available from the corresponding authors on reasonable request.

\subsection*{Supplementary materials}
Supplementary Text\\

\newpage
\renewcommand{\thefigure}{S\arabic{figure}}
\renewcommand{\thetable}{S\arabic{table}}
\renewcommand{\theequation}{S\arabic{equation}}
\renewcommand{\thepage}{S\arabic{page}}
\setcounter{figure}{0}
\setcounter{table}{0}
\setcounter{equation}{0}
\setcounter{page}{1}

%%%%%%%%%%%%%%%% SUPPLEMENT TITLE PAGE %%%%%%%%%%%%%%%

\begin{center}
\section*{Supplementary Materials for\\ \scititle}

	Pranshoo Upadhyay$^{1,\ast}$,
	Daniel G. Su\'arez-Forero$^{1,2,\ast,\dagger}$,
	Tsung-Sheng Huang$^{1,\ast}$, \and
    Mahmoud Jalali Mehrabad$^{1}$,
    Beini Gao$^{1}$,
    Supratik Sarkar$^{1}$,
    Deric Session$^{1}$, \and
    Kenji Watanabe$^{3}$,
    Takashi Taniguchi$^{3}$,
    You Zhou$^{4,5}$, \and
    Michael Knap$^{6,7}$,
    Mohammad Hafezi$^{1,\ddagger}$\and\\
	\small$^{1}$Joint Quantum Institute, University of Maryland, College Park, MD 20742, USA.\\
    \small$^{2}$Department of Quantum Matter Physics, University of Geneva, \\Geneva, Switzerland.\\
    \small$^{3}$Research Center for Materials Nanoarchitectonics, National Institute for Materials Science,\\ 1-1 Namiki, Tsukuba 305-0044, Japan.\\
    \small$^{4}$Department of Materials Science and Engineering, University of Maryland,\\ College Park, MD 20742, USA.\\
    \small$^{5}$Maryland Quantum Materials Center, College Park, Maryland 20742, USA.\\
    \small$^{6}$Technical University of Munich, TUM School of Natural Sciences, \\Physics Department, 85748 Garching, Germany.\\
    \small$^{7}$Munich Center for Quantum Science and Technology (MCQST), \\Schellingstr. 4, 80799 M{\"u}nchen, Germany.\\
	\small Corresponding authors. Email: $^\dagger$ suarez@umbc.edu, $^\ddagger$ hafezi@umd.edu \\
	\small$^\ast$These authors contributed equally to this work.
\end{center}

\subsubsection*{This PDF file includes:}
Supplementary Note 1. Device structure and optical characterization \\
Supplementary Note 2. Formation of AP and its dependence on pump intensity \\
Supplementary Note 3. Ultra-low power excitation regime \\
Supplementary Note 4. Diffusion in extended doping range in R- and H-stacked devices\\
Supplementary Note 5. Characteristic diffusion length in steady-state condition\\
Supplementary Note 6. Chromatic aberration check and raw spatial PL profile\\
Supplementary Note 7. Spectrally resolved diffusion \\
Supplementary Note 8. Reproducing data in a different H-stacked device\\
Supplementary Note 9. Doping-dependent renormalization of moir\'e potential \\
Supplementary Note 10. Temperature dependence of Mott gap 
\clearpage
\section*{Supplementary Note 1. Device structure and optical characterization}
\begin{figure}[h]
    \centering
    \includegraphics[width=1\linewidth]{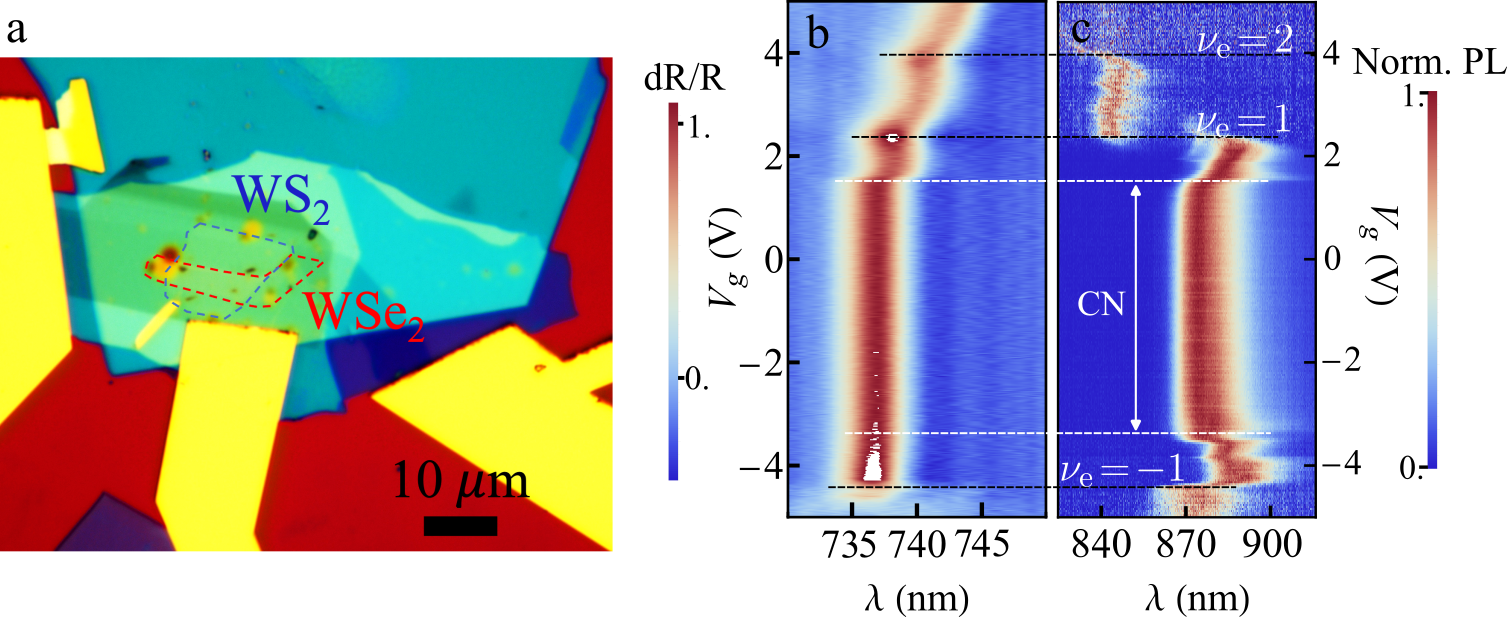}
    \caption{\textbf{WS$_2$/WSe$_2$ heterobilayer and its optical characterization.} a) The optical micrograph of the heterobilayer device. WS$_2$ and WSe$_2$ monolayer regions are indicated by the blue and red dashed lines, respectively. b) Electronic filling ($\nu_{\rm{e}}$)-dependent reflection contrast spectrum of WSe$_2$ intralayer exciton on the bilayer region indicates the Mott-insulating and band-insulating states at $\nu_{\rm{e}}$ = 1 and 2, respectively. c) PL spectrum of IX features AP for electron and hole doping and doublon-hole pair (X2) beyond $\nu_{\rm{e}}$ = $\pm$ 1. }
    \label{figS1}
\end{figure}

We fabricate a transition metal dichalcogenide (TMD) hetero-bilayer sample for diffusion measurements. WS$_2$ and WSe$_2$ are the two monolayer TMDs used for this purpose. After aligning their edges, they are stacked to achieve 0$^{\rm{o}}$ twist angle, i.e., 2H stacking order. The sample is encapsulated with hBN ($\sim$ 35 nm) and gated on both sides for independent control of the out-of-plane electric field and doping. Figure \ref{figS1}a shows an optical micrograph of the sample. We characterize the device using reflection and photoluminescence (PL) measurements. Figures \ref{figS1}b and c present the reflection contrast spectrum of WSe$_2$ intralayer exciton and PL map of the interlayer exciton (IX) measured on the bilayer region, respectively. 
Both measurements show distinctive features at the charge-neutral (CN) region ($\nu_{\rm{e}}$ = 0), the Mott-insulating region ($\nu_{\rm{e}}$ = 1), and the band-insulating region ($\nu_{\rm{e}}$ = 2). Identifying these regions is useful to calibrate the gate voltage ($V_g$) to $\nu_{\rm{e}}$. However, for better accuracy, we track the peak intensity variations of the IX species with doping as shown in Fig.~2a (Main text). Here, $\nu_{\rm{e}}\!=\!0$ is benchmarked by the emergence of the peak associated with the attractive polaron (AP). The presence of that peak coincides with a decrease in the X1 intensity. $\nu_{\rm{e}}\!=\!1$ is identified by the complete extinction of the X1 intensity. It is important to mention that a proper calibration of $\nu_{e}$ can only be performed at ultra-low pump intensities (a few nW/$\mu m^2$ or lower), because it relies on the features of the polaron gap, only present at these low exciton densities.

\section*{Supplementary Note 2. Formation of AP and its dependence on pump intensity}
\begin{figure}
    \centering
    \includegraphics[width=0.6\linewidth]{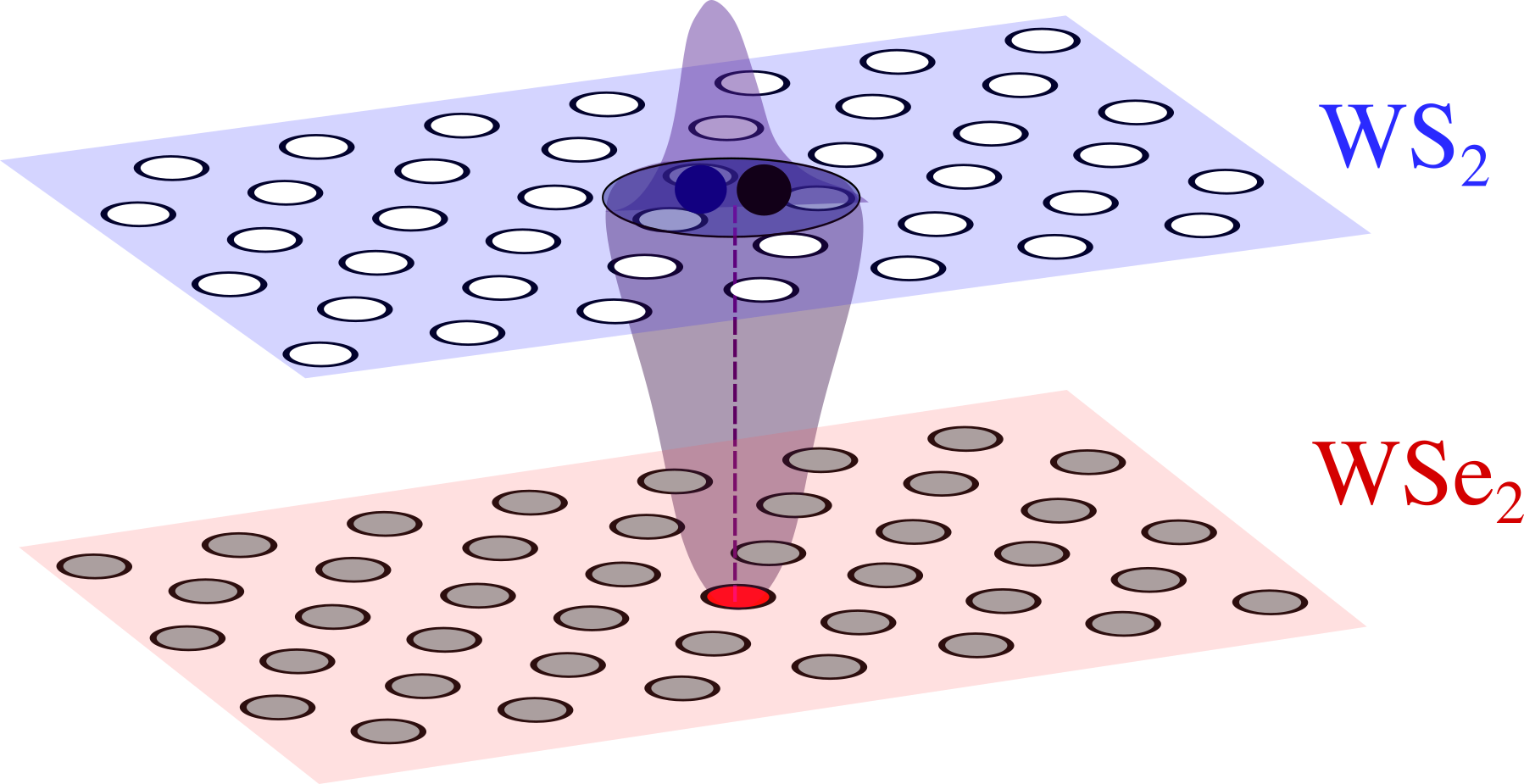}
    \caption{\textbf{Formation of interlayer AP in a moir\'e lattice.} Schematic of the AP formation in the H-stacked WS$_2$/WSe$_2$ heterostructure. Here, an exemplary case of X1 dressed by a doped electron is considered. As the hole resides in a different moir\'e registry than electrons, it has three electron sites with equal probability of binding. This results in the wavefunction of both the electrons spread across the three moir\'e sites.
    }
    \label{figS2}
\end{figure}

Due to the H-stacking order in this system, electrons and holes localize in different moir\'e registries as discussed in detail in recent literature \cite{ElectronspreadH2023}. This lateral separation of electrons and holes is the key to the formation of interlayer AP in the system. In this configuration, a hole in the WSe$_2$ layer can bind with three sites in the electron layer with equal probability. This allows the hole in the bottom layer (WSe$_2$) to pair with a cloud of electrons in the top layer (WS$_2$). Due to the complexity involved in understanding the precise nature of APs, specifically in the moir\'e system, we consider a trion-like charge complex, without affecting the conclusion of this work, to explain our observations. In Fig. \ref{figS2}, we illustrate this exemplary case of AP where an embedded X1 binds with one doped electron. 
\begin{figure}
    \centering
    \includegraphics[width=0.8\linewidth]{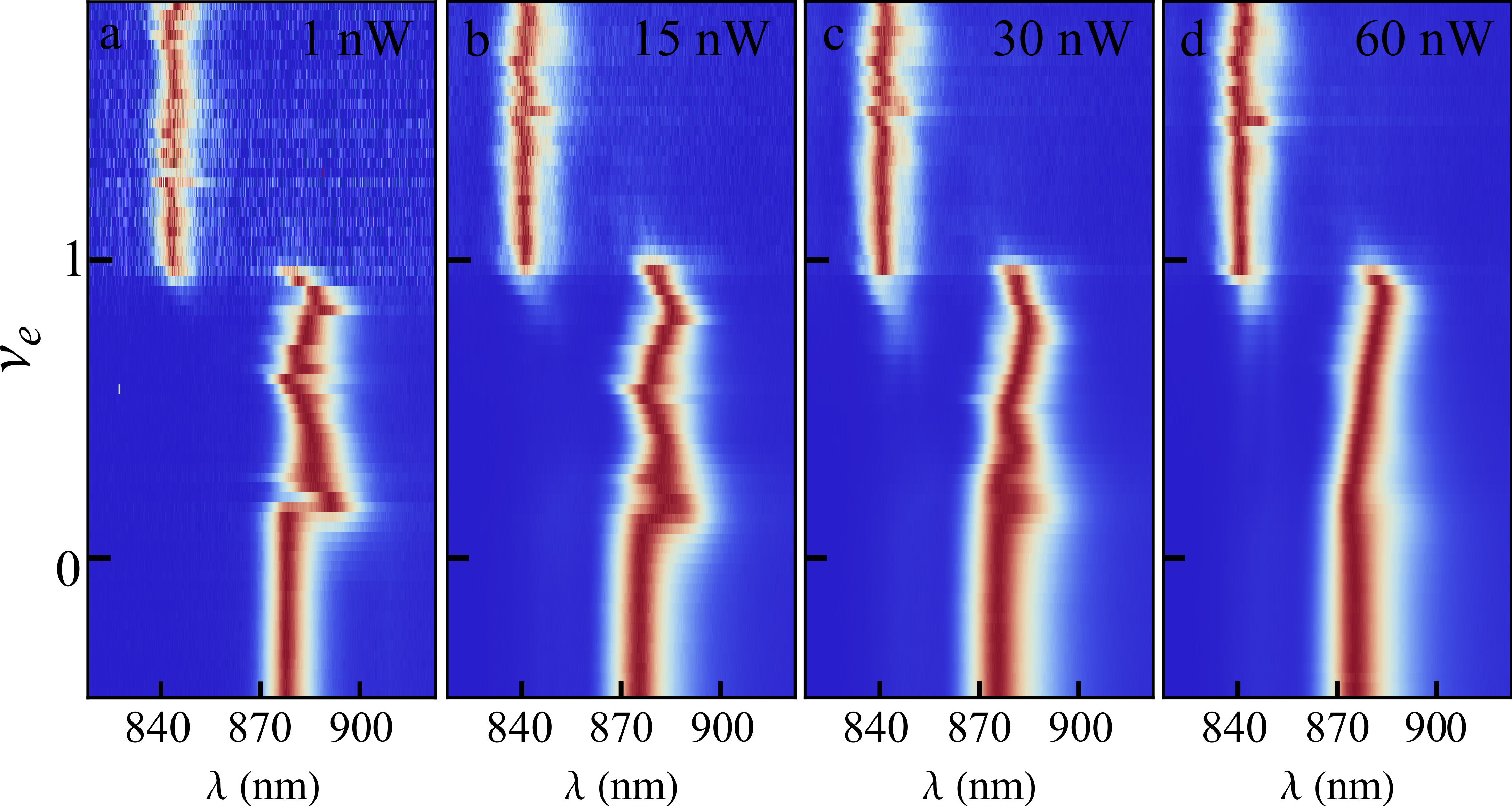}
    \caption{\textbf{Variation of polaron gap with excitation power.} The gap above the CN region gradually closes with the pump power. This confirms the presence of AP only in the ultra-low exciton density regime.}
    \label{figS3}
\end{figure}

It should be further noted that AP formation can be distinctively observed in PL only for ultra-low excitation power. When the density of excitons is lower than the density of doped electrons in the system, we can have each exciton associated with at least one electron, forming AP. However, here, the density of doped electrons must be small as at higher densities close to $\nu_{\rm{e}} \sim$ 1 the repulsive interactions become strong and inhibit the formation of AP. We illustrate the dependence of AP on exciton and electron density in Fig. \ref{figS3}. For the case of 1 nW/$\mu \rm{m}^2$, we have a clear transition from X1 to AP upon doping. This gap is observed to reduce and eventually vanish with excitation pump intensity. Additionally, AP transitions to X1 again at high electron doping close to $\nu_{\rm{e}} \sim$ 1.

\section*{Supplementary Note 3. Ultra-low power excitation regime}
\begin{figure}[h]
    \centering
    \includegraphics[width=0.5\linewidth]{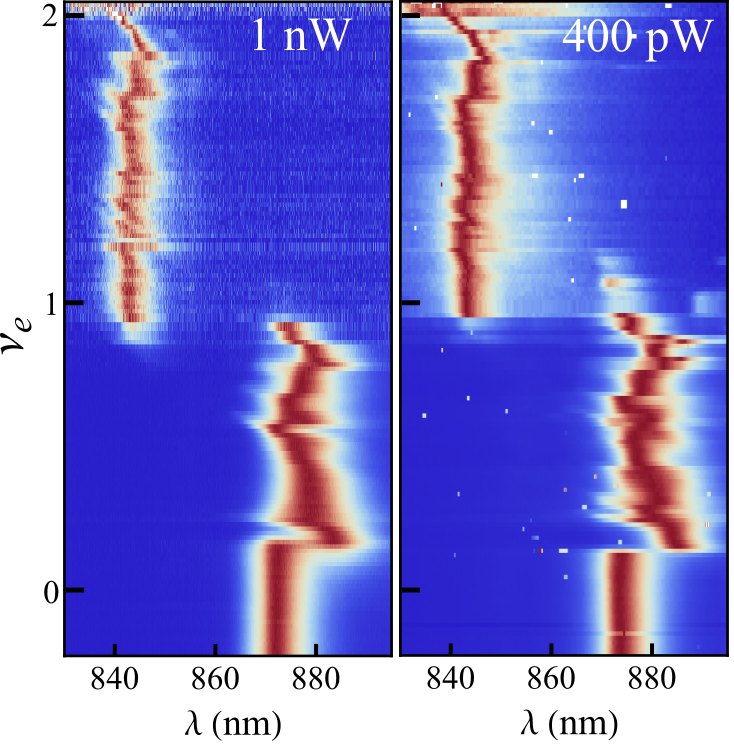}
    \caption{\textbf{PL of ultra-low exciton density.} PL at both 1nW/$\mu \rm{m}^2$ and 400 pW/$\mu \rm{m}^2$ have the same polaron gap and Mott transition conditions. This confirms the ultra-low exciton population regime.}
    \label{figS4}
\end{figure}

Based on the above discussion regarding AP formation, we determine the ultra-low exciton density regime. A complete polaron gap ($\sim$ 16 meV) is used as a benchmark for achieving this regime. In the main text, we have shown a complete polaron gap for 5 nW/$\mu \rm{m}^2$. In Fig. \ref{figS4}, we demonstrate that any pump intensity below 5 nW/$\mu \rm{m}^2$ will have a comparable spectrum in the entire range of doping. This confirms our regime of operation.

\section*{Supplementary Note 4. Diffusion in extended doping range in R- and H-stacked devices}
\begin{figure}[h]
    \centering
    \includegraphics[width=1\linewidth]{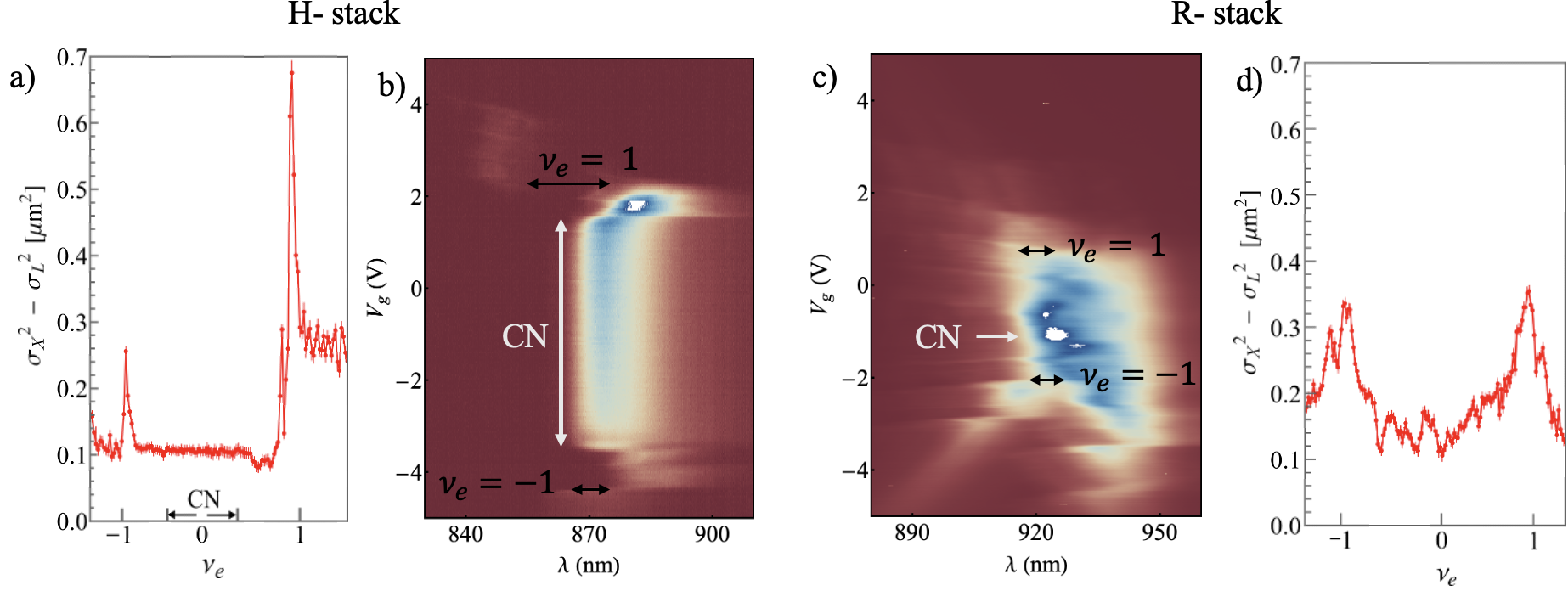}
    \caption{\textbf{Diffusion in the extended doping range.} a) Spatial diffusion and b) Spectrally resolved PL measured for  H-stack device. The diffusion peak is observed in both the electron and hole-doped regimes. The latter is weaker due to a much smaller Mott gap. c) PL and d) spatial diffusion in an R-stacked device agrees with the observation that Mott gap size is correlated with the strength of the diffusion peak.}
    \label{figSRandH}
\end{figure}

Here, we present the spatial diffusion modifications in both the hole- and electron-doped regimes. As shown in Fig. \ref{figSRandH}a, the diffusion peaks are observed near both the Mott insulating states ($\nu_e \sim -1$ and $\nu_e \sim 1$). However, the diffusion peak on the hole-doped side is much weaker. From the PL data, shown in Fig. \ref{figSRandH}b, the Mott gap at the electron-doped side is $\sim 6$ times larger than that of the hole-doped side. The Mott gap characterizes the strength of the exciton-charge Coulomb interactions.

Additionally, we perform diffusion measurements on R-stacked devices showing smaller Mott gaps (Fig. \ref{figSRandH}c) and hence weaker diffusion enhancement (Fig. \ref{figSRandH}d) compared to the electron-doped Mott insulating case in H-stacked devices. With these observations, we provide evidence that the diffusion peak near the Mott insulating state depends on the size of the Mott gap.

If the Mott gap is comparable to the inhomogeneous broadening, some regions may favor X2 formation (monogamous diffusion), preventing the sharp or giant diffusion enhancement observed in the electron-doped H-stack case. Remarkably, this is also corroborated with our temperature-dependent results where the diffusion peak vanishes (Fig. 4b) at temperatures below melting of the Mott state (Fig. S9).

\section*{Supplementary Note 5. Characteristic diffusion length in steady-state condition}

To accurately account for the diffusion of excitons under a CW excitation, we consider a time-independent (steady-state) driven-dissipative diffusion equation in 2D with a spatially localized source:
\begin{equation}
-D \nabla^2 n(\mathbf{r}) = R(\mathbf{r})-  \frac{n(\mathbf{r})}{\tau} ,
\end{equation}
where $D$ is the diffusion coefficient, $\tau$ is the exciton lifetime. The source is spatially localized Gaussian charaterized by:
    \begin{equation}
    R(\mathbf{r}) = \frac{R_0}{\pi \sigma_L^2} e^{-r^2/\sigma_L^2},
    \end{equation}
with $R_0$ and $\sigma_L$ denoting the intensity and width of the source, respectively.
Taking the 2D Fourier transform of $\tilde{n}(\mathbf{k}) = \int d^2r \, e^{-i\mathbf{k}\cdot\mathbf{r}} n(\mathbf{r})$, we obtain
\begin{equation}
\tilde{n}(\mathbf{k}) = \frac{R_0 e^{-k^2 \sigma_L^2 / 4}}{D k^2 + \frac{1}{\tau}}
.
\end{equation}
To find $n(\mathbf{r})=n(x, y)$, with $(x,y)$ denoting components of $\mathbf{r}$, we need to take the inverse Fourier transform $n(x, y) = \frac{1}{(2\pi)^2} \int d^2k \, e^{i\mathbf{k}\cdot\mathbf{r}} \tilde{n}(k_x, k_y)$, where $(k_x, k_y)$ are components of $\mathbf{k}$. Our objective is to calculate the diffusion length, $\sqrt{\langle r^2 \rangle} = \sqrt{\frac{\int dx \, dy \, (x^2 + y^2)n(x, y)}{\int dx \, dy \, n(x, y)}}$. We note that only $k = 0$ is required to evaluate $\sqrt{\langle r^2 \rangle}$ as:
\begin{equation}
\int dx \, dy \, n(x, y) = \tilde{n}(0, 0) = {R_0}\tau,
\end{equation}
\begin{equation}
\int dx \, dy \, (x^2 + y^2) n(x, y) = -\left(\left.\frac{\partial^2 \tilde{n}}{\partial k_x^2}\right|_{k=0} + \left.\frac{\partial^2 \tilde{n}}{\partial k_y^2}\right|_{k=0} \right)
.
\end{equation}
Utilizing the series expansion $\tilde{n}(k) = R_0\tau \left(1 - \left(D\tau + \frac{\sigma_L^2}{4} \right)k^2 + \cdots \right)$, we obtain:
\begin{equation}
\int dx \, dy \, (x^2 + y^2) n(x, y) = 4 \cdot R_0\tau \left(D\tau + \frac{\sigma_L^2}{4} \right),
\end{equation}
\begin{equation}
\langle r^2 \rangle = \frac{\int dx \, dy \, (x^2 + y^2)n(x, y)}{\int dx \, dy \, n(x, y)} = \sigma_L^2 + 4 D\tau
.
\end{equation}
Finally, our calculation demonstrates the characteristic diffusion length, $L_{\rm{d}}\equiv\sqrt{D\tau}$, to be proportional to $\sqrt{\langle r^2 \rangle -\sigma_L^2}$, which has been used in our analysis.

\section*{Supplementary Note 6. Chromatic aberration analysis and raw spatial PL profile}
\begin{figure}[h]
    \centering
    \includegraphics[width=1\linewidth]{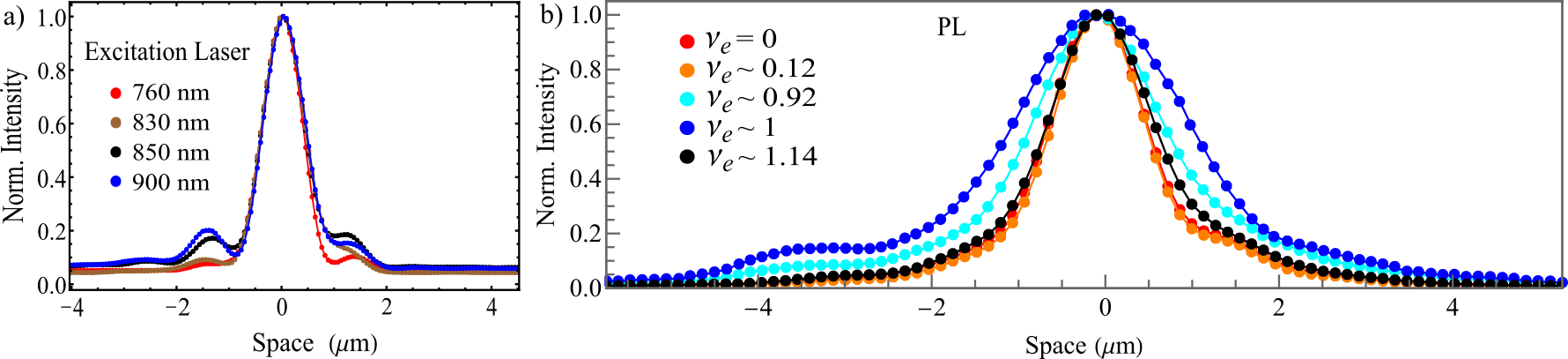}
    \caption{\textbf{Validating the Setup and observation.} a) Spatial profile of the excitation laser at different wavelengths. The minimal deviation in the collected size of the excitation spot at different excitation wavelengths indicates negligible chromatic aberration effects in the setup. b) The raw spatial profile of the collected PL in different doping regimes. In the Mott regime (blue markers), the broader Gaussian profile indicates enhanced diffusion.}
    \label{figSRaw}
\end{figure}

The PL response as a function of electron filling is rich in this bilayer heterostructure with three excitonic species in the range $0 \lesssim \nu_e \lesssim1$. Moreover, the PL energy variation in this range is significant. Therefore, testing the setup for any chromatic aberrations is important, as they can lead to artifacts in the spatial PL profile. In Fig. \ref{figSRaw}a, we show the laser spatial profile at different excitation wavelengths covering the entire range of PL emission in this device. The minimal variation observed in the spot size with excitation wavelengths confirms the negligible effect of chromatic aberration on the optical setup. After this careful consideration, we obtain the evident increase in spatial diffusion of excitonic species near the Mott insulating state as shown in Fig. \ref{figSRaw}b.

\begin{figure}[h]
    \centering
    \includegraphics[width=0.7\linewidth]{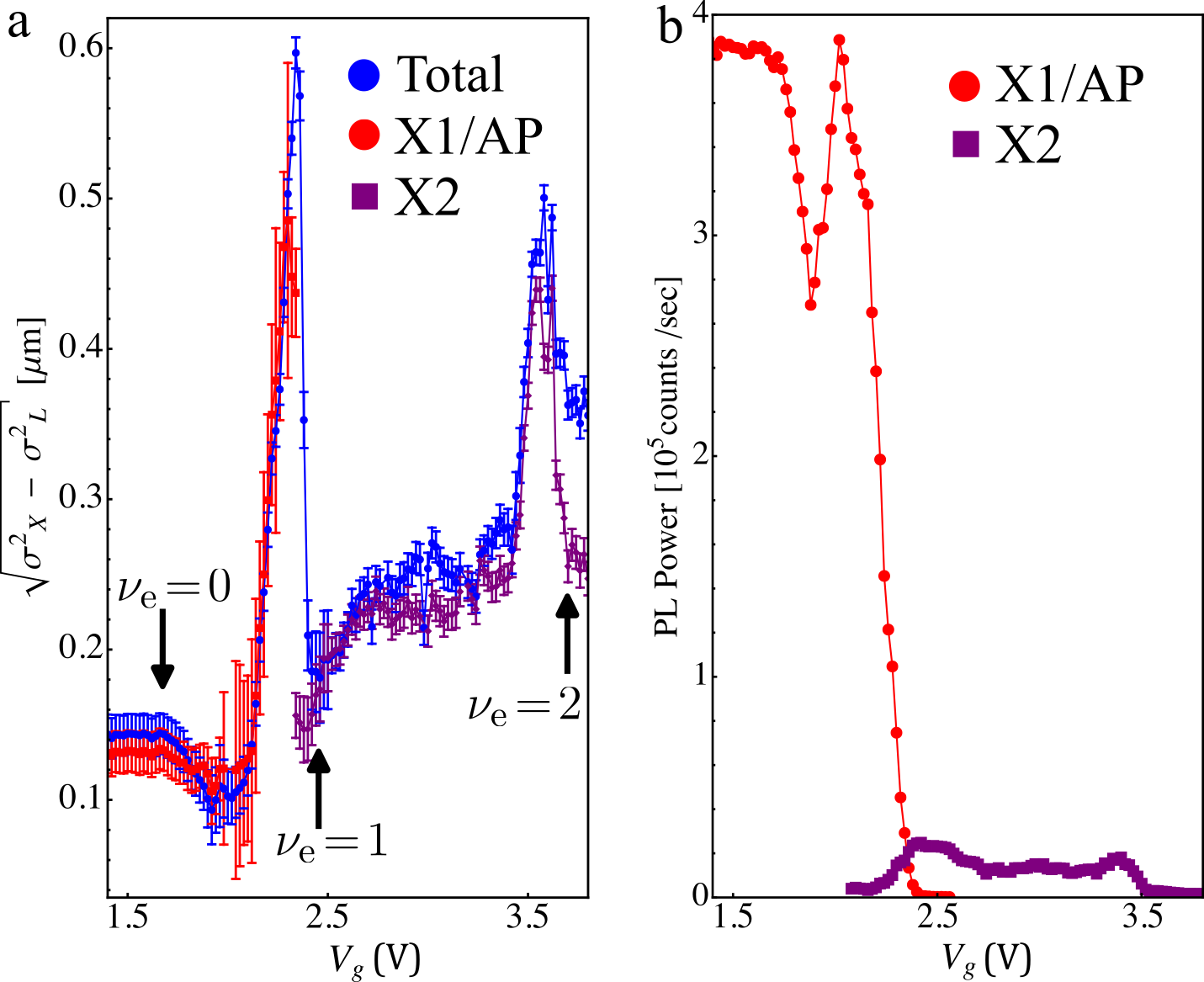}
    \caption{\textbf{Spectrally resolved diffusion.} a) Blue, red, and purple markers indicate the net diffusion (no spectral resolution), X1/AP diffusion, and X2 diffusion, respectively. Filling factors are marked with arrows. b) Integrated PL variation of X1/AP and X2 over the doping range indicates similar X1 and X2 intensity near the Mott region.}
    \label{figS5}
\end{figure}
\section*{Supplementary Note 7. Spectrally resolved diffusion}
As discussed in the main manuscript, the increase in diffusion length ($L_{\rm{d}}$) near the Mott-insulating region is a novel observation that warrants a detailed analysis. In Fig. \ref{figS5}a, we show spectrally resolved diffusion collected by using gratings and separating the spectrum into two regions: one below 860 nm (X2) and one above (X1 or AP). We apply the Gaussian fitting routine and extract the diffusion length. As shown by the red (X1 or AP) and purple (X2) markers, diffusion near the Mott region is solely from X1. Although both X1 and X2 have similar intensity in that region (Fig. \ref{figS5}b), X2 demonstrates negligible diffusion. However, for $\nu_{\rm{e}} \geq 1$, the diffusion is dominated by X2 and results in a peak near $\nu_{\rm{e}}$ = 2.

\section*{Supplementary Note 8. Reproducing data in a different H-stacked device}
\begin{figure}[h]
    \centering
    \includegraphics[width=0.7\linewidth]{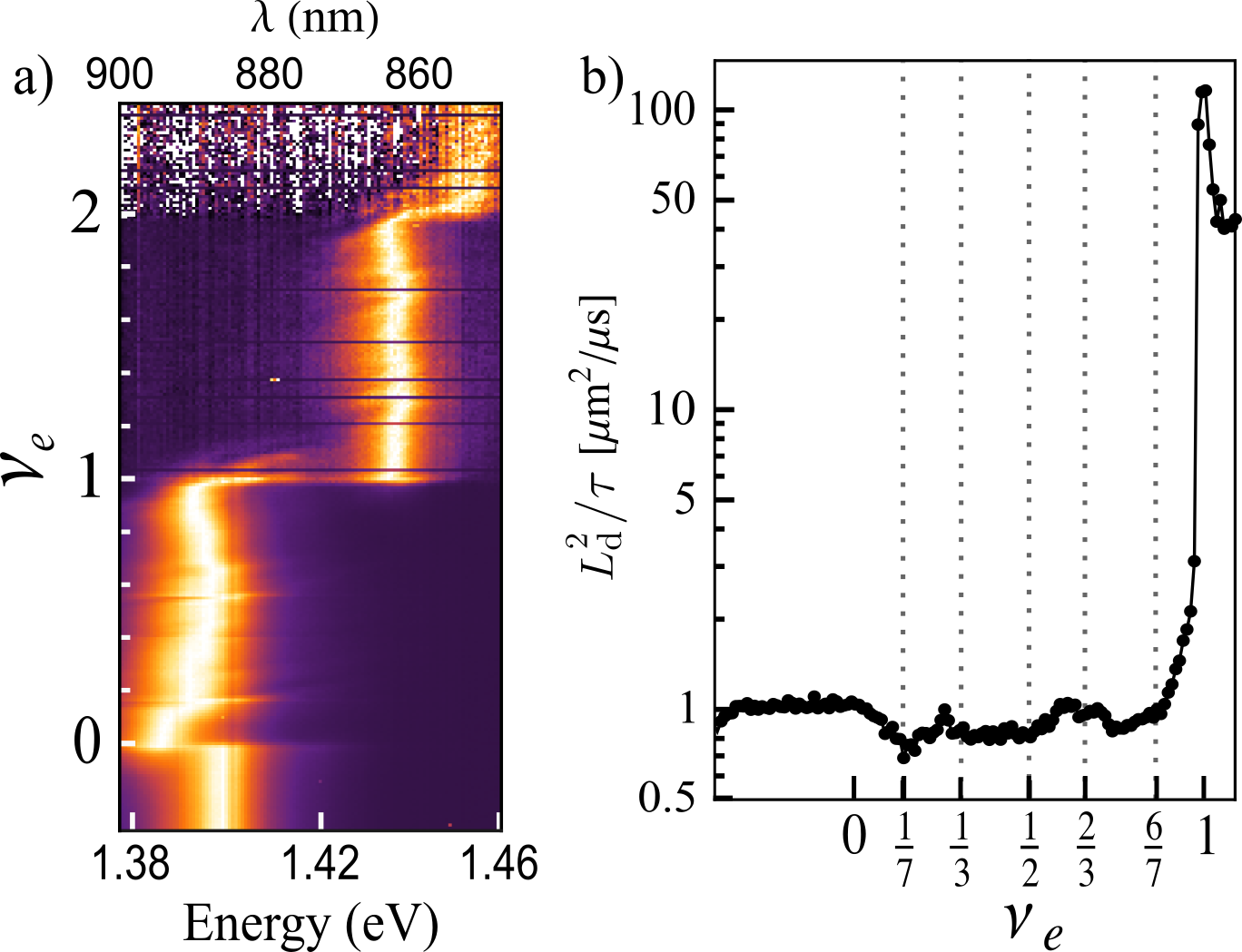}
    \caption{\textbf{PL and Diffusivity in another H-stacked device.} a) Spectrally resolved PL showing variation with electronic filling. b) Diffusivity as a function of electronic filling factor. Similar features are reproduced in this device as compared to Fig. \ref{fig2}d.}
    \label{figSH2}
\end{figure}

To confirm the enhancement in diffusivity near the Mott insulating state, we perform measurements on four different H-stacked devices. Here, we present a device that was fabricated and measured in a different facility. As shown in Fig. \ref{figSH2}a, the device shows a similar PL response as in Fig. \ref{fig1}f. After extracting the diffusion length and lifetime independently, the obtained diffusivity demonstrates similar enhancement near the Mott insulating state (Fig.~\ref{figSH2}b). Additionally, the modifications in $D$ in the presence of AP and fractional states, discussed in the main manuscript, are reproduced in this device.

\section*{Supplementary Note 9. Doping-dependent renormalization of moir\'e potential}

To model the variations in diffusivity with doping, we solve a simplified one-dimensional lattice Hamiltonian:
$\hat{H}=-\frac{\hbar^2}{2m}\frac{d^2}{d z^2}+\frac{V}{2} \sin\left(\frac{2\pi z}{\Lambda}\right).$
Here, $m$ is the mass of the exciton, trion, or hole --- depending on the regime of interest --- in the absence of the moiré potential. 
$\Lambda$ denotes the moiré lattice spacing, and $V$ is the effective superlattice potential renormalized by electrostatic interactions between the moving quasiparticle and the background electrons.
This Hamiltonian, whose eigenstates in real space are the Mathieu functions, has been successfully used to model interlayer excitons in TMD heterobilayers \cite{lagoin2021key}.
Notably, although this model does not capture the geometric details of the moiré lattice, it provides estimates of effective masses of the moving quasiparticles renormalized by the effective superlattice potential, $m^*$, which is crucial for understanding excitonic diffusion (as it is inversely proportional to the mobility). In the following, we use the values of $m$ and $V$ in each regime to determine the corresponding $m^*$, which enables us to interpret the diffusive behavior shown in Fig.~\ref{fig2}d of the main text. These values are summarized in Table~\ref{table}.

We begin our analysis with the charge-neutral (CN) regime, where the exciton moves monogamously. In this regime, with no background electrons, $V$ is the bare moir\'e potential $V_{\rm{m}}$, which we take to be $\sim$100 meV in each layer \cite{yuan2020twist}. Considering $m = m_{\rm{e}} + m_{\rm{h}}$, where $m_{\rm{e}} =$ 0.5 $m_{\rm{0}}$ and $m_{\rm{h}} =$ 0.4 $m_{\rm{0}}$ ($m_{\rm{0}}$ is the free electron mass) are the electron and hole mass, respectively \cite{amin2014strain}, we find $m^\ast \simeq$ 182 $m_{\rm{0}}$ for the undoped regime.

In the moderately-doped regime, all the charges bind together as a composite object (trion or polaron) such that there are effectively no background electrons. The major difference is that $m$ changes as the quasiparticle now involves more charges. Here, we focus on a trion-like situation where an excited X1 binds to one doped electron, as developing a full theoretical model for the attractive polaron in the presence of a lattice remains an open problem. Neglecting rotational motion (which is always true in this 1D picture), $m$ is simply the sum of the bare masses of all charges $2m_{\rm{e}} + m_{\rm{h}}$. Taking $V$ to be the same as in the CN region for simplicity, we find a much larger effective mass $m^\ast \simeq$ 2800 $m_{\rm{0}}$ (and consequently slower diffusion than X1) originating from the increase of $m$.

In contrast, background electrons are present near the Mott regime, and these electrons crystallize due to strong electron-electron repulsion.
In such a configuration, the electrostatic potential from the ordered background electrons acts as an additional superlattice potential to any extra charges. Consequently, the effective potential experienced by the moving quasiparticle, the hole, is $V = V_{\rm{m}} - V_{\rm{eh}}$, with the calculated $V_{\rm{eh}}\sim 88$ meV denoting the attraction between the background electrons and the hole. Notably, the effective moir\'e potential is reduced.
This reduction arises because the supersites of the electrons and the hole are arranged out of phase in H-stacked bilayer, indicating that the electrostatic attraction facilitates the hole's ability to cross its moir\'e potential barriers. The above considerations, together with $m = m_{\rm{h}}$, yield $m^\ast \simeq$ 0.4$m_{\rm{0}}$, $\sim$ 3 orders of magnitude reduced from that in the CN regime.

A similar discussion applies to the X2 regime, except that the moving quasiparticle is now a composite object formed by a hole and an ``extra'' electron. Here, the effective moir\'e potential of the hole is still $V = V_{\rm{m}} - V_{\rm{eh}}$; in contrast, the potential for the extra electron is renormalized by electron-electron repulsion, estimated to be $V_{\rm{ee}}\simeq$ 46 meV from PL data in Fig.~\ref{fig1}f, resulting in $V = V_{\rm{m}} - V_{\rm{ee}}$. Together with $m = m_{\rm{e}} + m_{\rm{h}}$, we find $m^\ast \simeq$ 4.6 $m_{\rm{0}}$, which is greater than in the Mott regime but orders of magnitude smaller than in the CN regime.

Finally, we note that in addition to $m^\ast$, the scattering rate between the moving quasiparticle and other species, $1/\tau_s$ (which is physically distinct from $1/\tau$), may also influence the diffusivity. However, we attribute the observed orders-of-magnitude increase in the diffusion coefficient near $\nu_{\rm{e}} = 1$ (see Fig. \ref{fig2}d) primarily to changes in $m^\ast$, rather than in $\tau_s$. This expectation is supported by the particle-hole symmetry between doped charge configurations at $\nu_{\rm{e}}$ and $1 - \nu_{\rm{e}}$, which implies that $\tau_s$ should be approximately symmetric about $\nu_{\rm{e}} = \frac{1}{2}$. However, the experimentally extracted $D$ is highly asymmetric about $\nu_{\rm{e}} = \frac{1}{2}$.Therefore, although scattering rate may affect $D$, the orders of magnitude increase in $D$ near the Mott insulating state indicates a stronger dependence on $m^\ast$. This asymmetry is captured in the effective mass variation from our simplistic model. Such an asymmetry prevails when including polaronic interaction effects while keeping the scattering time fixed, as theoretically shown in a recent work \cite{pichler2025purely}.

\begin{table}
    \centering
\begin{tabular}{|c|c|c|c|c|}
\hline
Fermionic Region (Diffusing particle) & $m$ ($m_{\rm{0}}$) & $V$ (meV) & $m^*$ ($m_{\rm{0}}$) \\
\hline
CN Region (X1) & 0.9 & 200 & 182 \\
\hline
Low doping (AP) & 1.4 & 200 & 2800 \\
\hline
Mott (hole) & 0.4 & 12 & 0.4 \\
\hline
High Doping (X2) & 0.9 & 66 & 4.6 \\
\hline
\end{tabular}
    \caption{Effective mass variation with doping}
    \label{table}
\end{table}

\section*{Supplementary Note 10. Temperature dependence of Mott gap}
\begin{figure}[h]
    \centering
    \includegraphics[width=0.95\linewidth]{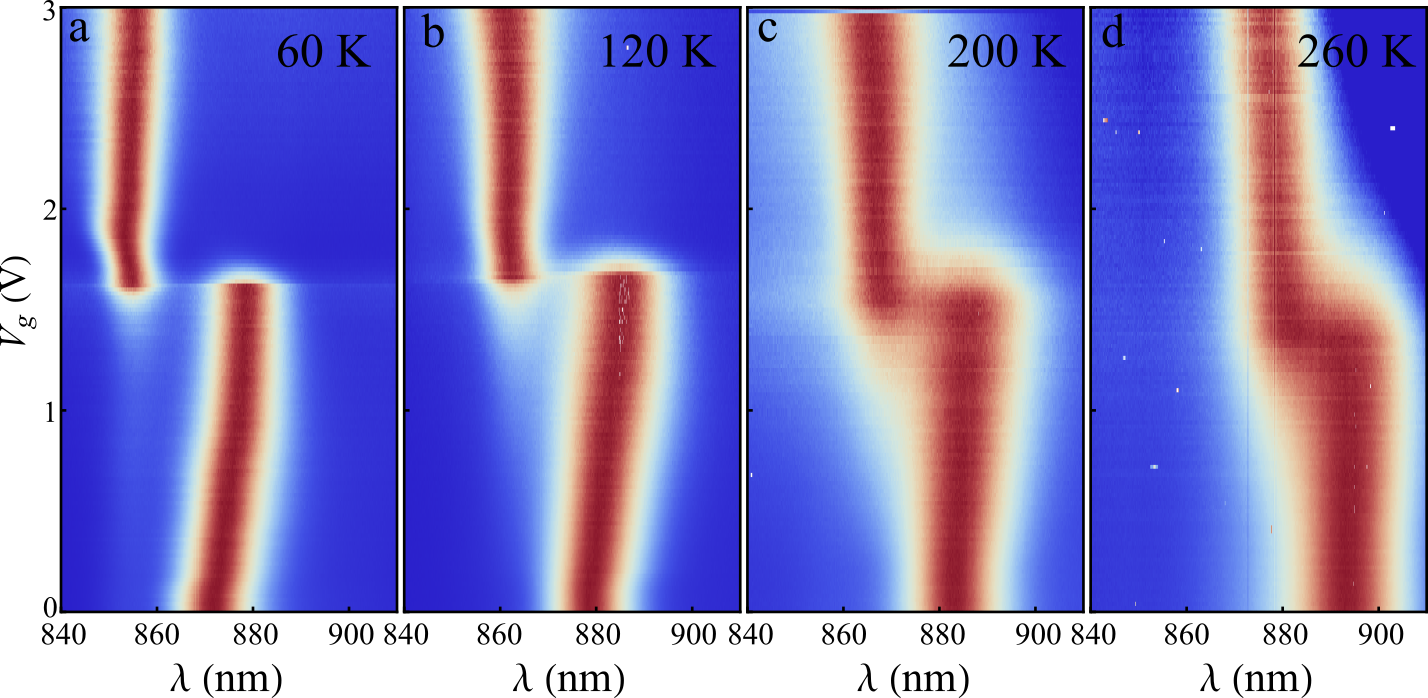}
    \caption{\textbf{Temperature dependence of Mott transition.} PL spectra at high temperatures, a) 60 K, b) 120 K, c) 200 K and d) 260 K demonstrate Mott gap. The gap, however, reduces with temperature as expected. Measurements are done at relatively higher pump intensity (300 nW/$\mu \rm{m}^2$).}
    \label{figS6}
\end{figure}

We perform temperature-dependent PL measurements to verify the presence of the Mott gap in the entire range of temperatures used for measurements. Figure \ref{figS6} demonstrates the Mott gap even at 260 K, although the gap reduction is observed with temperature.

\begin{comment}

\end{comment}
\end{document}